%% file: main.tex
\documentclass{bmcart}

\usepackage{amsthm,amsmath}
\usepackage[utf8]{inputenc} 

\def\includegraphics{}

\startlocaldefs
\endlocaldefs

\usepackage{graphicx}
\usepackage{sidecap}
\sidecaptionvpos{figure}{t}

\usepackage{lineno}
\modulolinenumbers[1]

\usepackage{multirow, makecell}
\usepackage{graphicx}
\usepackage{booktabs}
\usepackage{hyperref}

\usepackage{subfig}

\usepackage[ruled,vlined]{algorithm2e}
\usepackage{amsmath}

\hypersetup{
      allcolors=blue,
    colorlinks=true,
    citecolor=blue,
    urlcolor=magenta,
}

\usepackage{subfiles}
\usepackage{framed}
\usepackage[export]{adjustbox}

\usepackage{subfig}

\begin{document}
\begin{frontmatter}

\begin{fmbox}
\dochead{ }


\title{Machine Learning Based Texture Analysis of Patella from X-Rays for Detecting Patellofemoral Osteoarthritis}


\author[
  addressref={aff1},                   
  corref={aff1},                       
  email={name.surname@oulu.fi}   
]{\inits{N.B.}\fnm{Neslihan} \snm{Bayramoglu}}
\author[
  addressref={aff1,aff2,aff3},
  email={miika.nieminen@oulu.fi}
]{\inits{M.T.N.}\fnm{Miika T.} \snm{Nieminen}}
\author[
  addressref={ aff1,aff2,aff3},
  email={simo.saarakkala@oulu.fi}
]{\inits{S.S.}\fnm{Simo} \snm{Saarakkala}}


\address[id=aff1]{
  \orgdiv{\textbf{Research Unit of Medical Imaging, Physics and Technology}},             
  \orgname{ \textbf{University of Oulu}},          
  \city{\textbf{Oulu}},                              
  \cny{\textbf{Finland}}                                    
}
\address[id=aff2]{%
  \orgdiv{Department of Diagnostic Radiology},
  \orgname{Oulu University Hospital},
  \city{Oulu},
  \cny{Finland}
}
\address[id=aff3]{%
  \orgdiv{Medical Research Center},
  \orgname{University of Oulu and Oulu University Hospital},
  \city{Oulu},
  \cny{Finland}
}



\end{fmbox}


\begin{abstractbox}

\subfile{abstract.tex}





\end{abstractbox}
%

\end{frontmatter}
\pagebreak
\section*{Introduction}
\subfile{introduction.tex}

\subfile{proposal.tex}

\subfile{experiments.tex}

\subfile{Conclusion.tex}


\newpage
\pagebreak
\begin{backmatter}

\section*{Acknowledgements}

Multicenter Osteoarthritis Study (MOST) Funding Acknowledgment. MOST is comprised of four cooperative grants (Felson – AG18820; Torner – AG18832, Lewis – AG18947, and Nevitt – AG19069) funded by the National Institutes of Health, a branch
of the Department of Health and Human Services, and conducted by MOST study investigators. This manuscript was prepared using MOST data and does not necessarily reflect the opinions or views of MOST investigators.

We would like to acknowledge the strategic funding of the University of Oulu, Infotech Oulu.

We gratefully acknowledge Claudia Lindner for providing the BoneFinder\textsuperscript{\textregistered} tool and lateral knee active shape model, Aleksei Tiulpin for providing an interface to BoneFinder to fully leverage multiple processors, and the support of NVIDIA Corporation with the donation of the Quadro P6000 GPU used in this research.

\section*{Funding}
Funding sources are not associated with the scientific contents of the study.

\section*{Competing interests}
The authors declare that they have no competing interests.

\section*{Authors' contributions}
 N.B. originated the idea of the study. N.B. performed
the experiments and took major part in writing of the manuscript. M.T.N, and S.S. supervised the project. All  authors participated in producing the final manuscript draft and approved the final submitted version.

\section*{Summary Table}
\begin{itemize}
\item We  present  the  first  study  that  evaluates  patellar  bone  texture for  detecting  patellofemoral osteoarthritis (PFOA).
\item We proposed a framework based on machine learning and performed extensive experiments to demonstrate the influence of texture features on the diagnostic performance.

\item Different region of interests (ROIs) from patellar region from lateral view radiographs (X-rays) were studied whether texture features can be used to distinguish the knees with definite radiographic PFOA from healthy knee radiographs.

\item We compared the performances of hand-crafted features, deep convolutional neural network features,  and   
clinical variables including age, sex, body mass index (BMI), the total Western Ontario and McMaster Universities ArthritisIndex (WOMAC) score, and Kellgren and Lawrence (KL) score of the tibiofemoral joint.

\item Finally,  we  propose  a  stacked  model  where  both patellar texture and clinical feature predictions are combined with a second level machine learning model - Gradient Boosting Machine (GBM).

\item Our  results  show  that  texture  features  of  patellar  bone  are different between knees with and without PFOA.

\item Good classification performance values indicate that analyzed texture features contain useful information of patellar bone structure, which seems to change in PFOA.

\item Patellar bone texture features may be used in the future as novel imaging biomarkers in OA diagnostics.

\end{itemize}


\bibliographystyle{bmc-mathphys} 
\bibliography{bibliography}      





\end{backmatter}

\pagebreak
\section*{Supplementary}

\subfile{supplementary.tex}


\end{document}

%% file: abstract.tex
\begin{abstract}
\parttitle{Objective}
To assess the ability of texture features for detecting radiographic patello\-femoral osteoarthritis (PFOA) from knee lateral view radiographs.

\parttitle{Design}
We used lateral view knee radiographs from The Multicenter Osteoarthritis Study (MOST) public use datasets (n = 5507 knees). 
Patellar region-of-interest (ROI) was automatically detected using landmark detection tool (BoneFinder), and subsequently, these anatomical landmarks were used to extract three different texture ROIs. 
Hand-crafted features, based on Local Binary Patterns (LBP), were then extracted to describe the patellar texture.
First, a machine learning model (Gradient Boosting Machine) was trained to detect radiographic PFOA from the LBP features.
Furthermore, we used end-to-end trained deep convolutional neural networks (CNNs) directly on the texture patches for detecting the PFOA. 
The proposed classification models were eventually compared with more conventional reference models that use clinical assessments and participant characteristics such as age, sex, body mass index (BMI), the total Western Ontario and McMaster Universities Arthritis Index (WOMAC) score, and tibiofemoral Kellgren–Lawrence (KL) grade.
Atlas-guided visual assessment of PFOA  status by expert readers
provided in the MOST public use datasets was used as a classification outcome for the models. 
Performance of prediction models was assessed using the area under the receiver operating characteristic curve (ROC AUC), the area under the precision-recall (PR) curve -average precision (AP)-, and Brier score in the stratified 5-fold cross validation setting.

\parttitle{Results}
Of the 5507 knees, 953 (17.3\%) had PFOA.
AUC and AP for the strongest reference model including age, sex, BMI, WOMAC score, and tibiofemoral KL grade to predict PFOA were 0.817 and 0.487, respectively.
Textural ROI classification using CNN significantly improved the prediction performance (ROC AUC= 0.889, AP= 0.714).

\parttitle{Conclusion}
We present the first study that analyses patellar bone texture for diagnosing PFOA. 
Our results demonstrates the potential of using texture features of patella to predict PFOA.
 
\end{abstract}

%% file: introduction.tex

\begin{figure}[!t]
\centerline{
\begin{adjustbox}{color = bmcblue, varwidth=1\textwidth,margin=3pt 1pt 3pt 5,frame=0.5pt }
\centering
\includegraphics[width=0.36\textwidth, trim=0.1cm 4cm 0cm 5.5cm, clip]{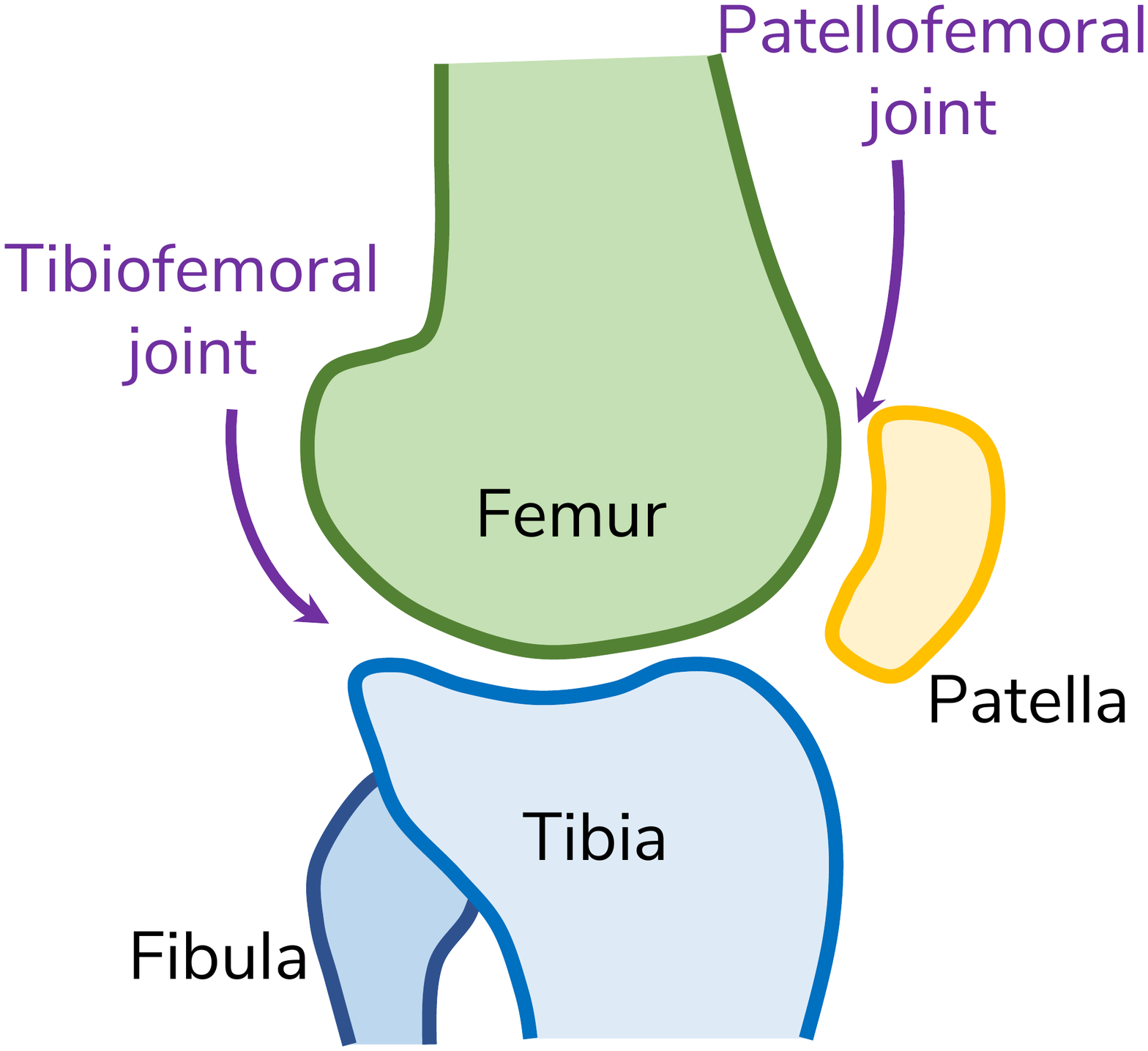}
\includegraphics[width = 0.6\linewidth, ]{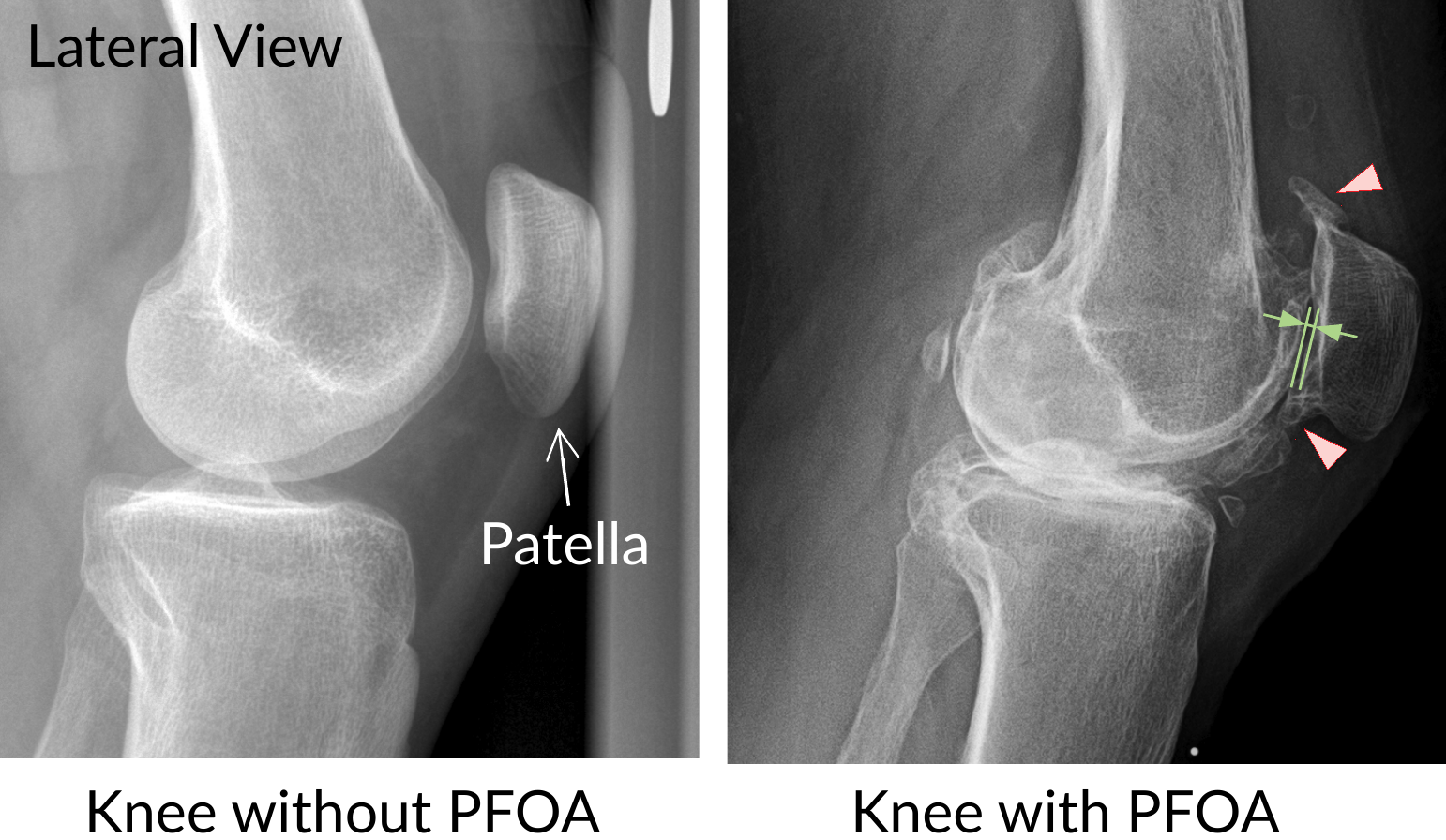}
\caption{ Figure on the left demonstrates the anatomy of the knee and knee joint articulations. Middle and right figure illustrate examples  of  lateral view X-ray  images  of normal knee and patellofemoral OA knee. 
On the right, joint  space  narrowing  (JSN) and osteophytes - characteristic features of OA - are demonstrated on the patellofemoral joint. }
\label{fig:sample}

\end{adjustbox}
}
\end{figure}

Osteoarthritis (OA) is the most common degenerative joint disease causing disability.
Knee is the most frequently affected of all joints.
Although the etiology of OA is not fully known yet, it is a complex disease resulting from combinations of risk factors including aging, female sex, obesity, past injury, and inflammation  \cite{bijlsma2011osteoarthritis}.
The disease has a profound effect on quality of life affecting both physical function and well-being.
Moreover, prevalence of knee OA is increasing due to aging population and increasing rates of obesity \cite{bijlsma2011osteoarthritis}.
Overall, the costs of knee OA are substantial.

There is no known cure for OA, and the disease is often progressive characterized with pain and loss of joint function.
Therefore, more research is needed to better understand the disease. Imaging biomarkers is one of the exciting areas that can advance our understanding.
Together with the developments in artificial intelligence, particularly deep learning, there is an increasing number of studies based on medical imaging of knee OA \cite{kokkotis2020machine}.
Most of these studies, however, focuses on tibiofemoral joint of the knee that consists of two condyloid articulation in which the femur has a rolling and sliding motion over the tibia \cite{kokkotis2020machine}.
However, there is a third articulation called  patellofemoral (PF) joint (Figure \ref{fig:sample}).
Osteoarthritis that occur at the PF joint, patellofemoral osteoarthritis (PFOA), is highly prevalent and clinically important yet still under-investigated \cite{hinman2007patellofemoral}.
For instance, there are clear diagnostic guidelines for tibiofemoral (TF) joint OA (\cite{altman2007atlas,kohn2016classifications,altman1986development,lee2020imaging}). Often these guidelines involve the use of well-validated radiographic grading systems for evaluating the severity of TFOA, such as  Kellgren and Lawrence (KL) grading system \cite{kellgren1957radiological} and
Osteoarthritis Research Society International (OARSI) standardized atlas \cite{altman2007atlas}.
However, there are no similar diagnostic guidelines explicitly developed for PF joint.
Moreover, in clinical setting, radiographic TF joint is often evaluated with a posterior-anterior (PA) view radiograph from which the PF joint can not be evaluated at all.
In addition, clinical assessments and participant characteristics alone cannot be used to diagnose patellofemoral OA \cite{tan2020can, stefanik2018diagnostic, bayramoglu2021pfoa}.
Since it has been reported that, at least in some phenotypes of knee OA, PF joint changes even precede TF joint changes (\cite{duncan2011incidence,stefanik2016changes,van2018international}), there is a strong need to consider also PF joint when developing new imaging biomarkers to diagnose and monitor knee OA .

In the OA research field, several studies have demonstrated the use of texture analysis for knee radiographs to distinguish knees with OA and without OA  \cite{bayramoglu2019adaptive, bayramoglu2020light, hirvasniemi2017differences, janvier2015roi}. However, these studies are all focusing on TFOA. With regard to PFOA, we showed recently that it can be detected using deep convolutional neural networks (CNNs) \cite{bayramoglu2021pfoa}.
However, such deep learning based models are ``black-boxes" and it is uncertain whether the model decisions are based on texture or some other imaging features such as shape or alignment. Therefore, in this study we investigated automated texture analysis of patella from lateral view radiographs. Specifically, we studied whether texture features can be used to distinguish the knees with definite radiographic PFOA from healthy knee radiographs. We proposed a framework based on machine learning and performed extensive experiments to demonstrate the influence of texture features on the diagnostic performance. Different region of interests (ROIs) from patellar region from lateral view radiographs (X-rays) were explored. Subsequently, we compared the performances of hand-crafted features, deep CNN features, and clinical features. Finally, we propose a stacked model where both patellar texture and clinical feature predictions are combined with a second level machine learning model - Gradient Boosting Machine (GBM) \cite{friedman2001greedy}. To the best of our knowledge, this is the first study to evaluate patellar bone texture in OA research.

\begin{figure}[!ht]
    \centerline{
    \begin{adjustbox}{color = bmcblue, varwidth=1.1\textwidth,margin=3pt 1pt 3pt 0,frame=0.5pt }
    \includegraphics[width=1\textwidth]{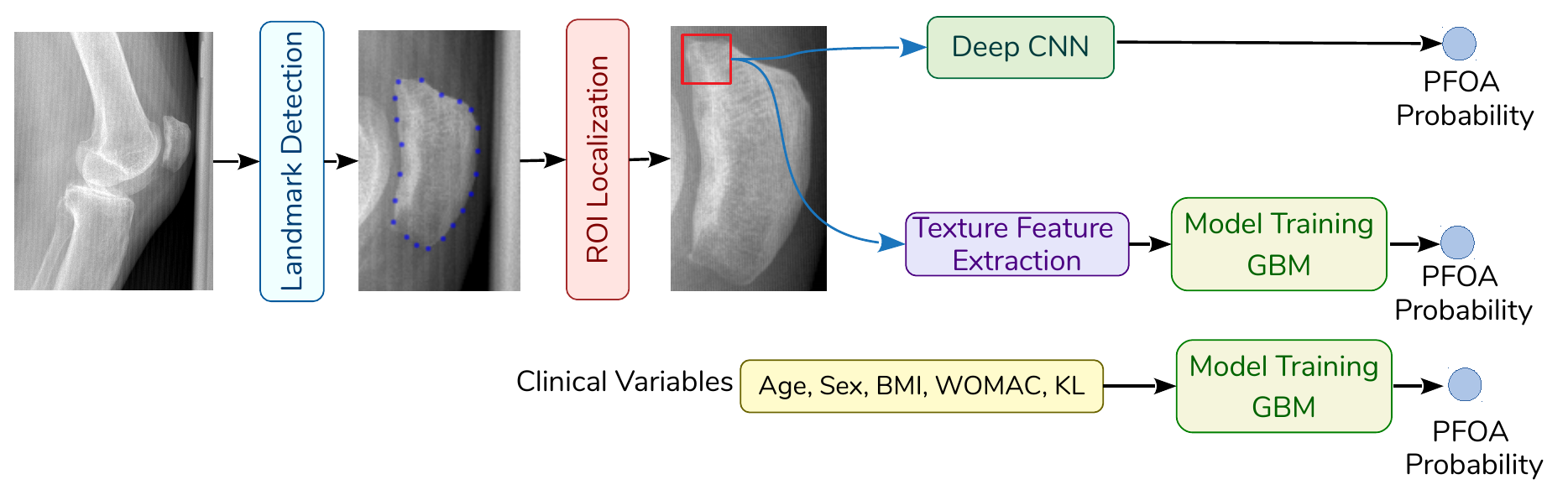}

    \caption{Illustration of the workflow of our approach.  First, we localized patellar landmarks using BoneFinder software (see Materials and Methods for  more  details).   Subsequently, we applied  intensity normalization and resampled the data to have a 0.2 mm pixel spacing.
Finally,  each  knee  is  rotated  in  order  to  have  an  aligned  patella.  
Three different regions of interest (ROIs) were located using patellar bone landmarks, after which a deep convolutional neural network (CNN) model was employed to predict the patellofemoral osteoarthritis (PFOA).
    Furthermore, we also trained a gradient boosting machine (GBM) model using handcrafted texture features (local binary patterns).
    To make a comparison of the proposed X-ray based methods, we trained another GBM model using clinical variables including age, sex, body mass index (BMI), the total Western Ontario and McMaster Universities Arthritis Index (WOMAC) score, and Kellgren and Lawrence (KL) score of the tibiofemoral joint.
    We used a stratified subject-wise 5-fold cross validation setting to measure the performance of all the methods.
    In addition to these individual models, we fused the predictions from these models in a second layer GBM model to  improve the overall prediction performance.
    }
    \label{fig:flowchart}
    
    \end{adjustbox}
    }
\end{figure}

%% file: proposal.tex
\section*{Materials and Methods}
The overall pipeline of our study is shown in Figure \ref{fig:flowchart}.
In order to pre-process the data, we extracted anatomical landmark points \cite{lindner2013fully} and applied intensity normalization using global contrast normalisation and a histogram truncation between the $5^{th}$ and $99^{th}$ percentiles.
We then resampled the data to have a 0.2mm pixel spacing.
Subsequently, we located ROIs using landmark points.
Right knee images were then horizontally flipped to have a similar view with left knee images.
Finally, we predicted PFOA using both handcrafted texture features using a machine learning-based approach and a deep CNN.
We also trained a machine learning model (GBM \cite{friedman2001greedy}) on clinical features as a reference  method  to  compare  with  the  proposed  approach.

\subsection*{\textbf{Data}}
In the study, we used data from the Multicenter Osteoarthritis Study public use datasets (MOST, \href{http://most.ucsf.edu}{http://most.ucsf.edu}).
The MOST study is a longitudinal observational study of adults who have or are at high risk
for knee OA.
At baseline, there were 3,026 individuals aged 50–79 years who either had radiographic knee OA or were at high risk for developing the disease.
In MOST, semiflexed lateral view radiographs were acquired according to a standardized protocol.
Knee radiographs were read from the baseline to 15, 30, 60 and 84-month follow-up visits.
We employed radiographs taken at the baseline visit that includes  5507 knees after removing data which have missing information. Out of 5507 knees, 953 had PFOA(17.3\%).

\begin{figure}[!ht]
\centering
\includegraphics[width = 1\linewidth, ]{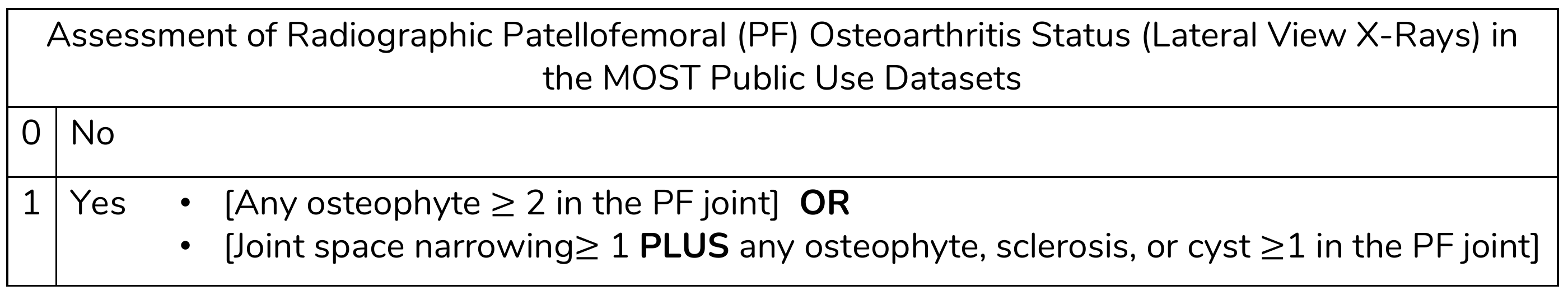}
\caption{Assessment of patellofemoral osteoarthritis in the MOST public use datasets. }
\label{fig:assessment}
\end{figure}

In the MOST public use datasets, radiographic PFOA is defined from lateral view radiographs as follows: Osteophyte score $\geq$ 2 or the joint space narrowing (JSN) score is $\geq1$ plus any osteophyte, sclerosis or cysts $\geq1$ in the PF joint (grades 0–3; 0=normal, 1=mild, 2=moderate, 3=severe) (Figure \ref{fig:assessment}).

Individual radiographic features in the MOST dataset were  graded by two independent expert readers based on the atlases from the Osteoarthritis Research Society International (OARSI) \cite{altman2007atlas} which refers to the previous OARSI atlas for the patellofemoral joint \cite{altman1995atlas} and Framingham Osteoarthritis Study \cite{chaisson2000detecting}.
When there was a disagreement in film readings, a panel of three adjudicators resolved the discrepancies\cite{roemer2009association}.

\subsection*{\textbf{Patellar Landmark Localization}}
\begin{figure}[!tb]

\centerline{
\begin{adjustbox}{color = bmcblue, varwidth=1\textwidth,margin=3pt 1pt 3pt 5,frame=0.5pt }
\centering
    \includegraphics[height=0.6\textwidth]{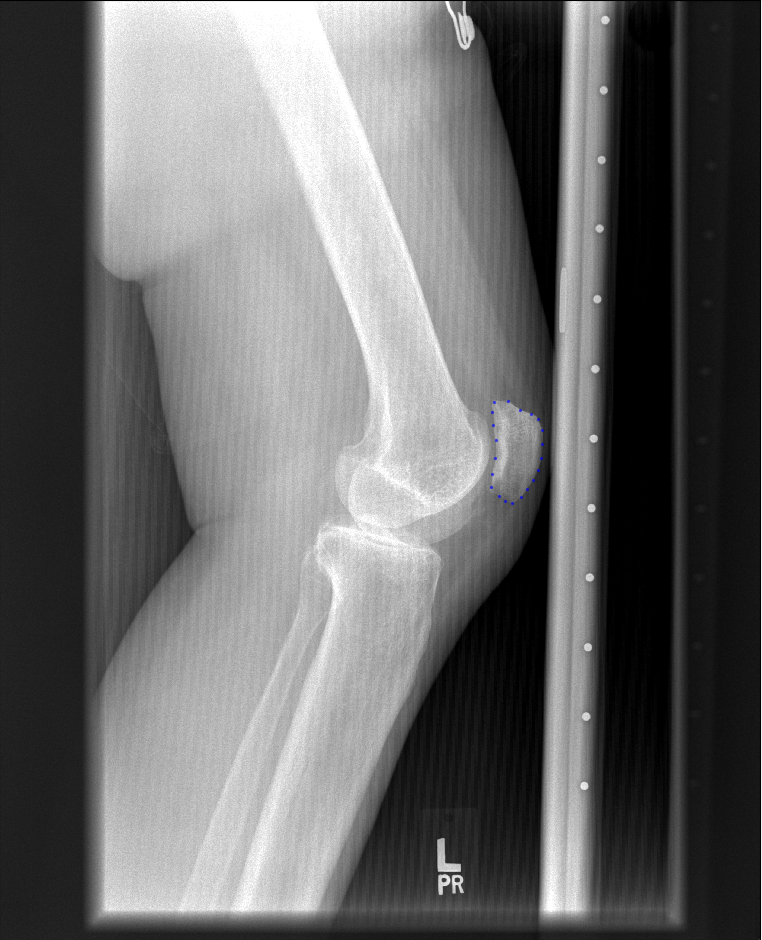}
    \includegraphics[height=0.6\textwidth]{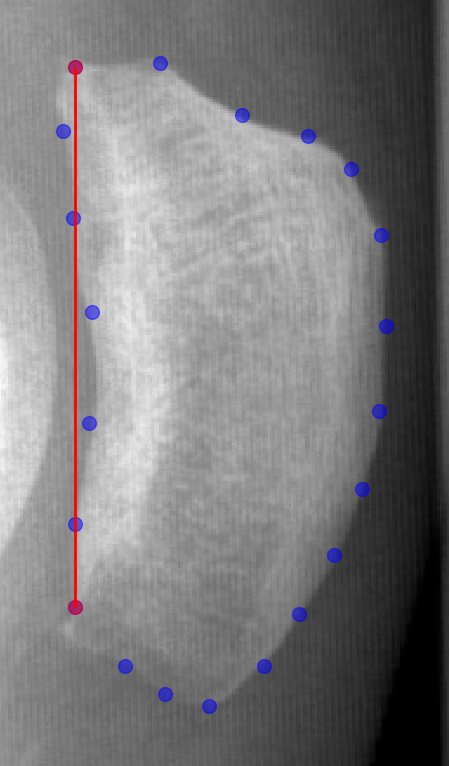}

    \centering
    \caption{Anatomical patellar landmarks were detected from lateral view knee radiographs. BoneFinder\textsuperscript{\textregistered} software \cite{lindner2013fully} were used to locate 21 landmark points along the contours of the patella.
    Marginal landmarks were employed in order to align the patellar region. The line on the right between the marginal points were draw in order to illustrate the final alignment of patella. }
    \label{fig:landmarks}
    
\end{adjustbox}
}
\end{figure}

In this study, we utilized BoneFinder\textsuperscript{\textregistered} software \cite{lindner2013fully} in order to locate the landmark points along the contours of the patella (Figure \ref{fig:landmarks}).
BoneFinder\textsuperscript{\textregistered} uses random forest regression voting with constrained local model approach to automatically locate 21 anatomical landmarks in patellar region from lateral knee X-rays.
These points enabled us to locate ROIs within the patella with precision.
Moreover, we used two marginal landmarks in order to align patella to obtain the same anatomical region among the knees in the ROI localization step (Figure \ref{fig:landmarks}).

\subsection*{\textbf{Texture Descriptor - Local Binary Patterns (LBP) }}
Previously in OA research, multiple texture descriptors have been applied to plain knee radiographs for several years,
Fractal Signature Analysis (FSA) or fractal dimension (FD) being the most widely used ones \cite{lynch1991analysis, janvier2017subchondral,hirvasniemi2017differences,kraus2018predictive, bayramoglu2019adaptive}.
However, FD method is susceptible to limited discrimination power due to its sensitivity to image artifacts and noise \cite{veenland1996estimation}.
It was shown in \cite{bayramoglu2019adaptive} that Local Binary Patterns (LBP)  \cite{ojala2002multiresolution} yields  better  performance in describing bone texture to detect OA for tibiofemoral joint.
Therefore, in our study, we selected LBP as a handcrafted texture descriptor as well.

LBPs captures a local representation of texture.
In order to compute LBP value of a pixel \textit{(c)}, \textit{p} neighboring pixels  that are evenly distributed in angle on a circle of radius \textit{r} are sampled first.
Then the LBP pattern is constructed by comparing the gray value of the center pixel (c) with its surrounding neighborhood of \textit{p} pixels and a binary vector of \textit{p} bits is extracted from this comparison.
The decimal value of LBP pattern is then obtained from the binary sequence.
For an $N\times M$ texture image, a LBP pattern can be the computed at each pixel -($N\times M$) LBP patterns-, then the image texture can be represented by the distribution of LBP values, by a LBP histogram vector.
In this study, we used $r=2$, $p=8\times r$ and a 256 bins histogram to obtain LBP descriptions of our texture patches.

\subsection*{\textbf{Selection of Region of Interest (ROI)}}
\begin{figure}[!h]
    \centerline{

\begin{adjustbox}{color = bmcblue, varwidth=1\textwidth,margin=3pt 1pt 3pt 5,frame=0.5pt }
\centering
    
    \subfloat[Superior ROI]{\includegraphics[height=0.4\textwidth]{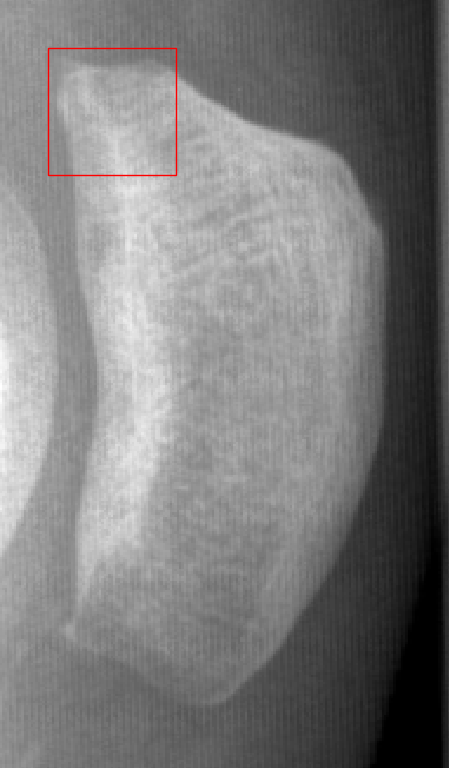}}
    \subfloat[Inferior ROI]{\includegraphics[height=0.4\textwidth]{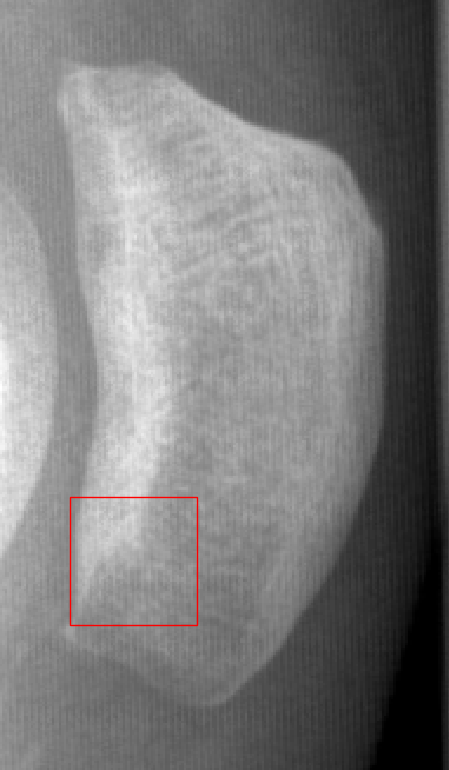}}
    \subfloat[Whole Patella]{\includegraphics[height=0.4\textwidth]{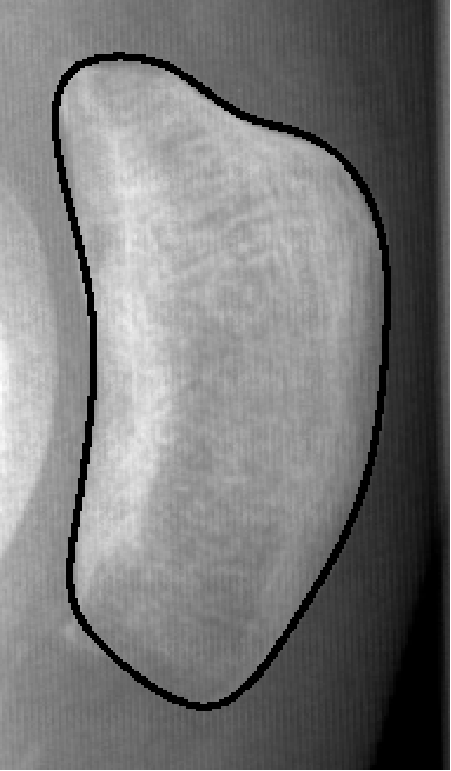}}

    \caption{ Examples  of  region-of-interest  (ROI)  placement  for  texture  analysis. Here,  a) shows an example of superior ROI, b) shows inferior ROI,  and  c)  whole patellar region. }

    \label{fig:patellar_roi}
    \end{adjustbox}
    }
\end{figure}

We explored three different ROIs within patellar region for texture analysis (Figure \ref{fig:patellar_roi}).
It is  known that along with the progression of OA, bone is subject to changes in its structure and composition \cite{buckland2004subchondral,kamibayashi1995trabecular}.
The radiographically most distinctive bony changes occur at the margins where osteophytes typically occur\cite{bayramoglu2019adaptive}.
Therefore, we selected two ROIs from the inferior and superior region of patella . 
Their sizes are proportional to the height of the patella measured from the outermost points ($20\%$ of the patella height).
To locate those ROIs, we used the marginal landmarks which were also utilized previously for alignment (Figure \ref{fig:landmarks}).
As the third ROI, we utilized the whole patellar region.
For segmenting patella from the background, we employed landmark points and obtained a smoothed spline that roughly approximates the contour  (Figure \ref{fig:patellar_roi}).

In order to detect the most informative (optimal) ROI, we employed LBP descriptor.
Here, we defined the most informative region as the subregion where the texture classifier (based on LBP features) performs best to distinguish OA samples from non-OA.

We used Gradient Boosting Machine (GBM) classifier based on decision tree algorithms \cite{friedman2001greedy} to predict PFOA using LBP features that were extracted from ROIs defined previously (Figure \ref{fig:patellar_roi}).
We observed that superior ROI features yields better classification performance (See Results section).
Therefore, we utilized superior ROI ( Figure \ref{fig:patellar_roi}a) in the subsequent experiments.

\subsection*{\textbf{Machine Learning Models}}
We employed both GBM and deep CNN methods to predict PFOA from the texture patches and clinical assessments and participant characteristics.
GBM is a powerful decision trees based machine learning algorithm that rely on the concept of boosting ``weak learners”\cite{friedman2001greedy}.
GBM is an iterative process - in each iteration training set is re-weighted such that it compensates for the weaknesses of the previous model.
In this study, we used an efficient implementation of GBM called LightGBM \cite{ke2017lightgbm}.
Table \ref{tab:comparison} summarises all the models developed in this study.

First, we extracted LBP texture features from superior ROI texture patches and trained a GBM model based on these features (Model1).
Then, we used a deep CNN architecture that was trained from end-to-end to classify superior ROI texture  (Model6).
In this way, texture features were learned from the imaging data itself.
CNNs captures both fine-level high spatial-frequency details such as edges, lines, texture, corners, etc and recognize more complex features as it gets deeper (as the number of convolutional layer increases).
We refer reader to \cite{Goodfellow-et-al-2016} and \cite{nielsen2015neural} for more information on the deep neural networks.
We used a three layers' CNN model where the input patches are scaled to $64\times64$ images.
Details of the CNN architecture and our training strategy can be found in the Supplementary Material.
In this study, we also explored more conventional machine learning based prediction models using the clinical data and risk factors.
These include age, sex, body-mass index (BMI), the total Western Ontario and McMaster Universities Arthritis Index (WOMAC) score, and the KL grade of the tibiofemoral joint (Model2,3,4,5).
These reference methods were also built using the GBM classifier.

In all models, we employed subject-wise stratified 5-fold cross validation.
Subject-wise splitting is used to eliminate the subject-dependent bias between training and validation.
That is, all the data (imaging and/or clinical) coming from a particular subject is either put in the training or the test set.
Moreover, we used stratified folds where each fold represents the actual class distribution of the data.
The class distribution is the ratio between the positive and negative samples.
The same folds were used for all the models to have fair comparisons.
All the models were trained separately, thus reported performances were derived from the separate models.

\subsection*{\textbf{Statistical Analyses}}

The models were assessed using Receiver Operating Characteristics (ROC) curves, Precision-Recall (PR) curves, and Brier score \cite{brier1950verification}.
In ROC curves, the true positive rate (TPR) against the false positive rate (FPR) are plotted, whereas Precision-Recall (PR) curves are composed of the precision and the true positive rate.
Therefore, PR analysis is more focused on the positive class. 
When data is imbalanced (i.e. more negative samples than positive samples), the ROC curve might not reflect the true performance of the classifier as false positive rate increases more slowly because of the large numbers of negatives.
We used the area under the  ROC  curves  (ROC  AUC) and similarly area under the PR curves (Average Precision; AP) to summarize model performances.
Brier score equals to the mean squared error of the prediction.
In order to compare the differences between model AUCs, we applied DeLong's test \cite{delong1988comparing}.

%% file: experiments.tex
\section*{Results}

\begin{table}[]
\centerline{
\begin{tabular}{@{}lllllll@{}}
\toprule
                            & Input                                                                                                               & Method                   & AUC                                           & AP                                            & \begin{tabular}[c]{@{}l@{}}Brier \\ Score\end{tabular} &                \\ \midrule
\multicolumn{1}{l|}{Model1} & \multicolumn{1}{l|}{\begin{tabular}[c]{@{}l@{}}LBP  Texture Features \\ (extracted from superior ROI)\end{tabular}} & \multicolumn{1}{l|}{GBM} & \multicolumn{1}{l|}{0.884 {[}0.871, 0.895{]}} & \multicolumn{1}{l|}{0.697 {[}0.669, 0.722{]}} & \multicolumn{1}{l|}{0.087}                             & Texture Model  \\ \midrule
\multicolumn{1}{l|}{Model2} & \multicolumn{1}{l|}{Age, Sex, BMI}                                                                                  & \multicolumn{1}{l|}{GBM} & \multicolumn{1}{l|}{0.647 {[}0.626, 0.666{]}} & \multicolumn{1}{l|}{0.294 {[}0.273, 0.317{]}} & \multicolumn{1}{l|}{0.136}                             & Clinical Model \\ \midrule
\multicolumn{1}{l|}{Model3} & \multicolumn{1}{l|}{Age, Sex, BMI, WOMAC}                                                                           & \multicolumn{1}{l|}{GBM} & \multicolumn{1}{l|}{0.715 {[}0.696, 0.732{]}} & \multicolumn{1}{l|}{0.351 {[}0.325, 0.378{]}} & \multicolumn{1}{l|}{0.130}                             & Clinical Model \\ \midrule
\multicolumn{1}{l|}{Model4} & \multicolumn{1}{l|}{Age, Sex, BMI, KL}                                                                              & \multicolumn{1}{l|}{GBM} & \multicolumn{1}{l|}{0.812 {[}0.798, 0.826{]}} & \multicolumn{1}{l|}{0.47 {[}0.441, 0.5{]}}    & \multicolumn{1}{l|}{0.114}                             & Clinical Model \\ \midrule
\multicolumn{1}{l|}{Model5} & \multicolumn{1}{l|}{Age, Sex, BMI, WOMAC, KL}                                                                       & \multicolumn{1}{l|}{GBM} & \multicolumn{1}{l|}{0.817 {[}0.802, 0.831{]}} & \multicolumn{1}{l|}{0.487 {[}0.457, 0.517{]}} & \multicolumn{1}{l|}{0.113}                             & Clinical Model \\ \midrule

\multicolumn{1}{l|}{Model6} & \multicolumn{1}{l|}{\begin{tabular}[c]{@{}l@{}}Superior ROI\\ (image patch)\end{tabular}}                           & \multicolumn{1}{l|}{CNN} & \multicolumn{1}{l|}{0.889 {[}0.875, 0.9}      & \multicolumn{1}{l|}{0.714 {[}0.687, 0.739{]}} & \multicolumn{1}{l|}{0.084}                             & CNN model      \\ \midrule

\multicolumn{1}{l|}{Model7}                     & \multicolumn{1}{l|} {\begin{tabular}[c]{@{}l@{}} LBP, Age, Sex, BMI \\WOMAC, KL\end{tabular} }                                      &  \multicolumn{1}{l|} {GBM}                      & \multicolumn{1}{l|}{{0.904 [0.892, 0.913]}   }          & \multicolumn{1}{l|}{{0.719 (0.693, 0.743)} }            & \multicolumn{1}{l|}{0.083}                                         & Fused Model  \\ \midrule

\multicolumn{1}{l|}{Model8}                     & \multicolumn{1}{l|} {\begin{tabular}[c]{@{}l@{}}Predictions from \\ Model5 and Model6\end{tabular} }                                      &  \multicolumn{1}{l|} {GBM}                      & \multicolumn{1}{l|}{\textbf{0.937 {[}0.929, 0.945{]}}   }          & \multicolumn{1}{l|}{\textbf{0.791 {[}0.768, 0.813{]}} }            & \multicolumn{1}{l|}{\textbf{0.069}}                                         & Stacked Model  \\ \bottomrule
\end{tabular}
}
\caption{Comparison of the developed models}
\label{tab:comparison}
\end{table}

\subsection*{Comparison of Region of Interest (ROI)}
\begin{figure} [!htb]
    \centerline{
    \includegraphics[width=0.6\textwidth]{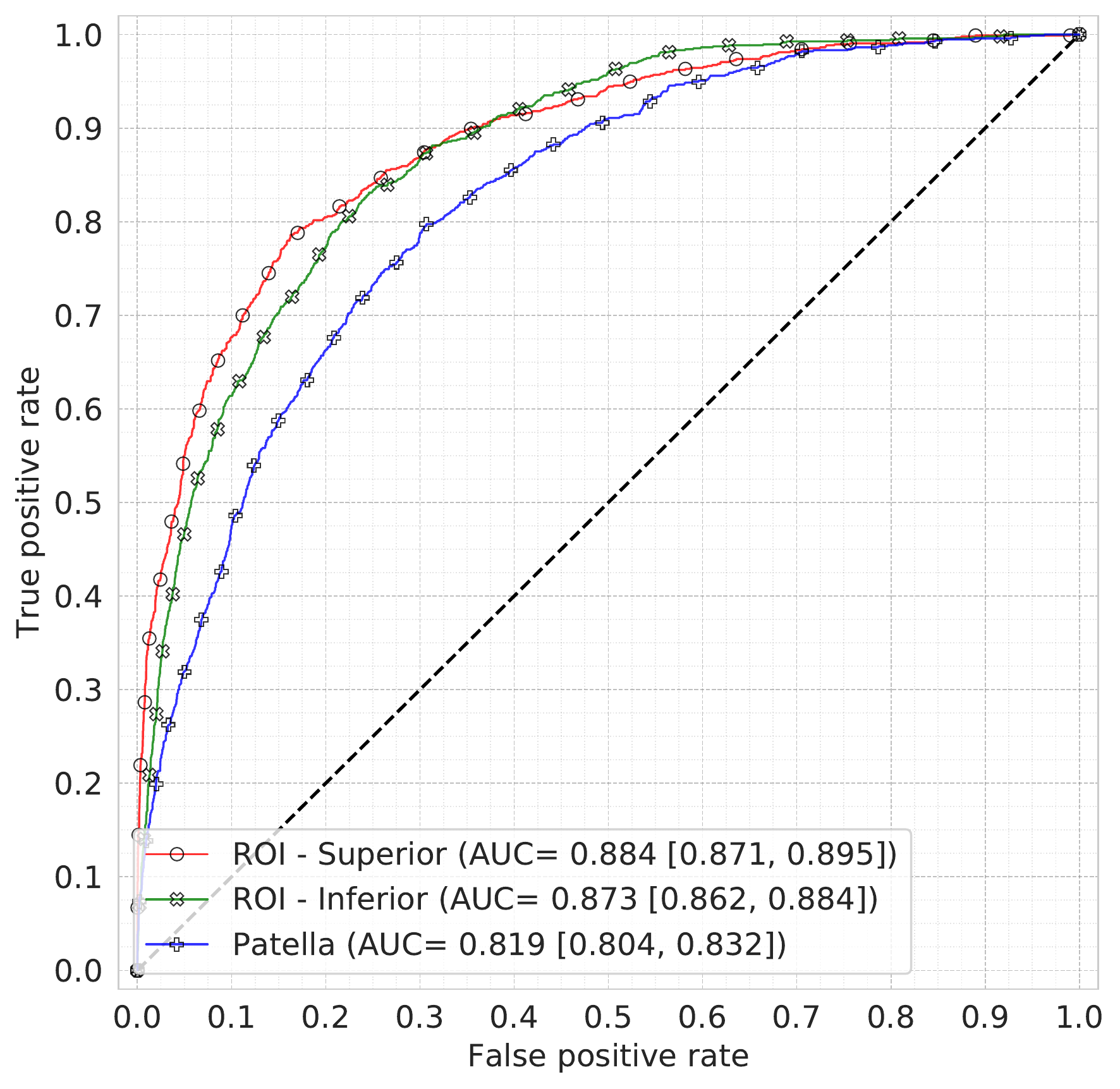}
    \includegraphics[width=0.6\textwidth]{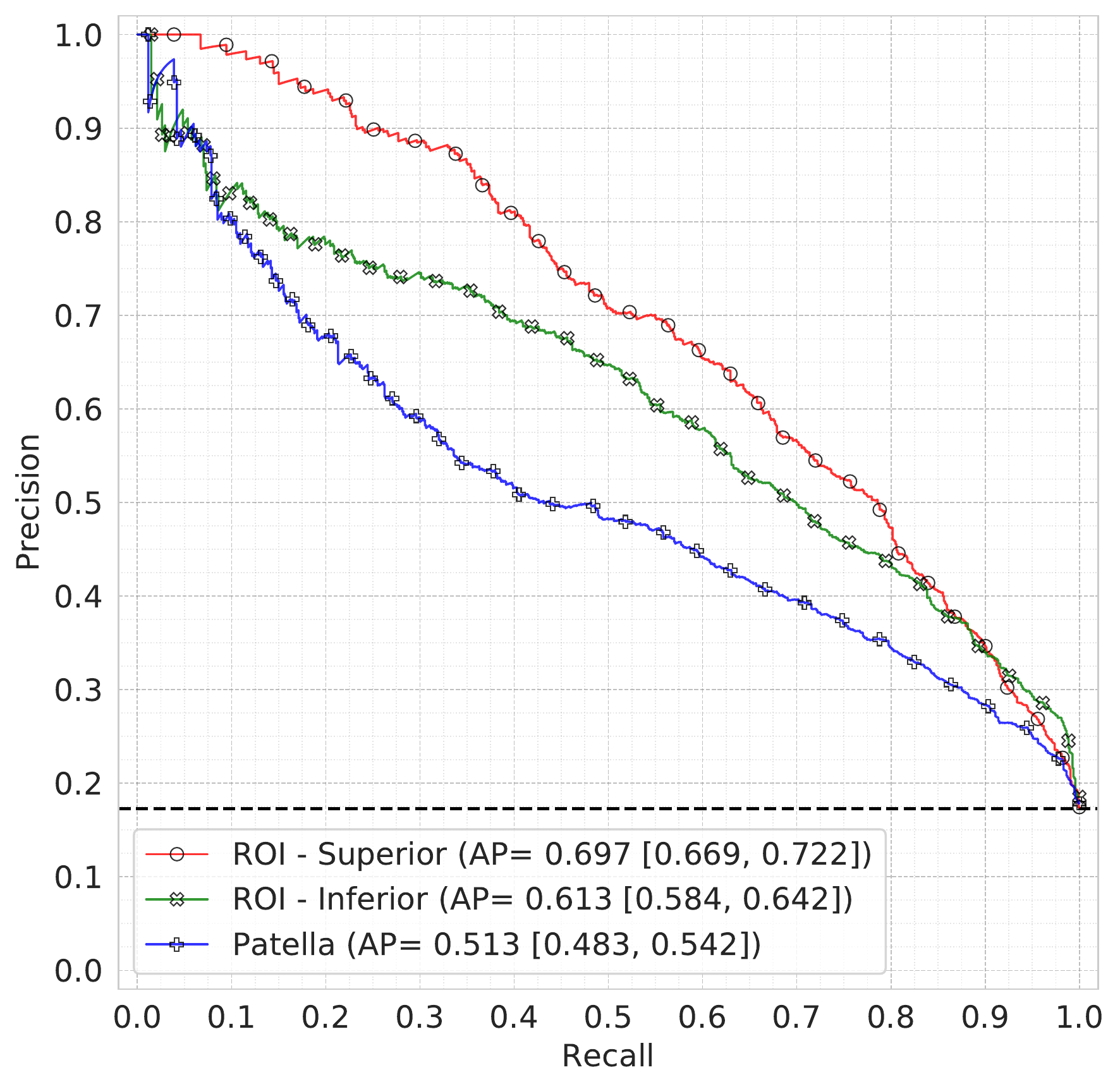}\hfill}
    \caption{ (a) ROC and (b) PR curves demonstrating the performance of the models operating on different region of interests (ROIs) - superior ROI, inferior ROI, and whole patella. 
    Area under the curves and 95\% confidence intervals in parentheses were given based on a 5-fold cross validation setting.
    Dashed lines in ROC indicate the performance of a random classifier and in case of PR it indicates the distributions of the labels of the dataset (PFOA vs non-PFOA).}
    \label{fig:roi_comparison}
\end{figure}

Firstly, we compared the performance of the texture descriptor (LBP) on the superior ROI, inferior ROI, and also on the whole patella in subject-wise stratified 5-fold cross validation setting (Figure \ref{fig:roi_comparison}).
Here, we used GBM models to predict PFOA.
From Figure \ref{fig:roi_comparison}, it can be seen that the model employing superior ROI features performs best yielding an AUC of 0.884 (0871-0.895) and AP of 0.697 (0.669-0.722).
We chose the superior ROI in our further comparisons because it performs with higher precision at most recall levels.
We observed a lower performance when we used texture features extracted from the whole patella.

\begin{figure}[!htb]
    \centerline{
    \includegraphics[width=0.6\textwidth]{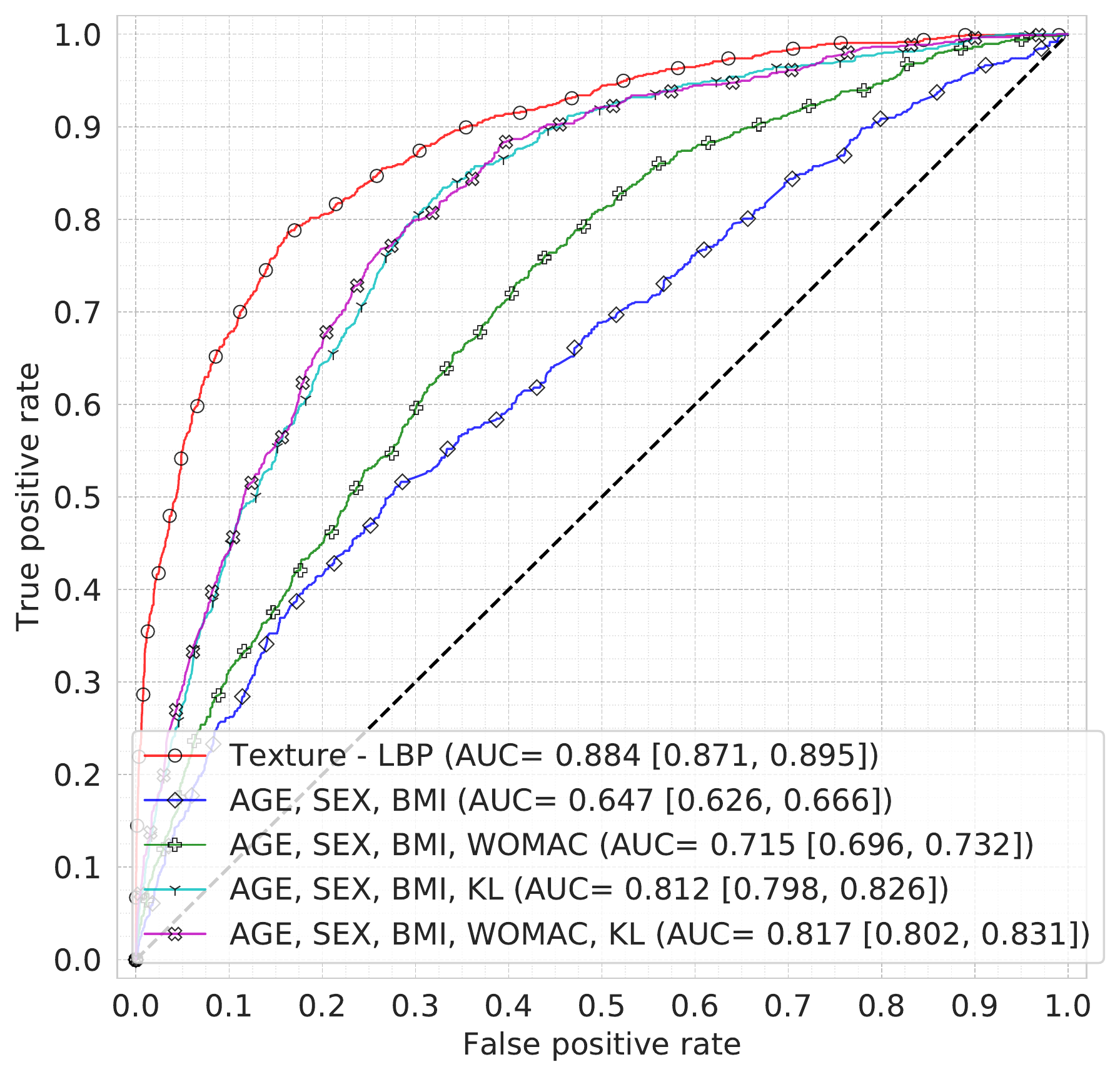}
    \includegraphics[width=0.6\textwidth]{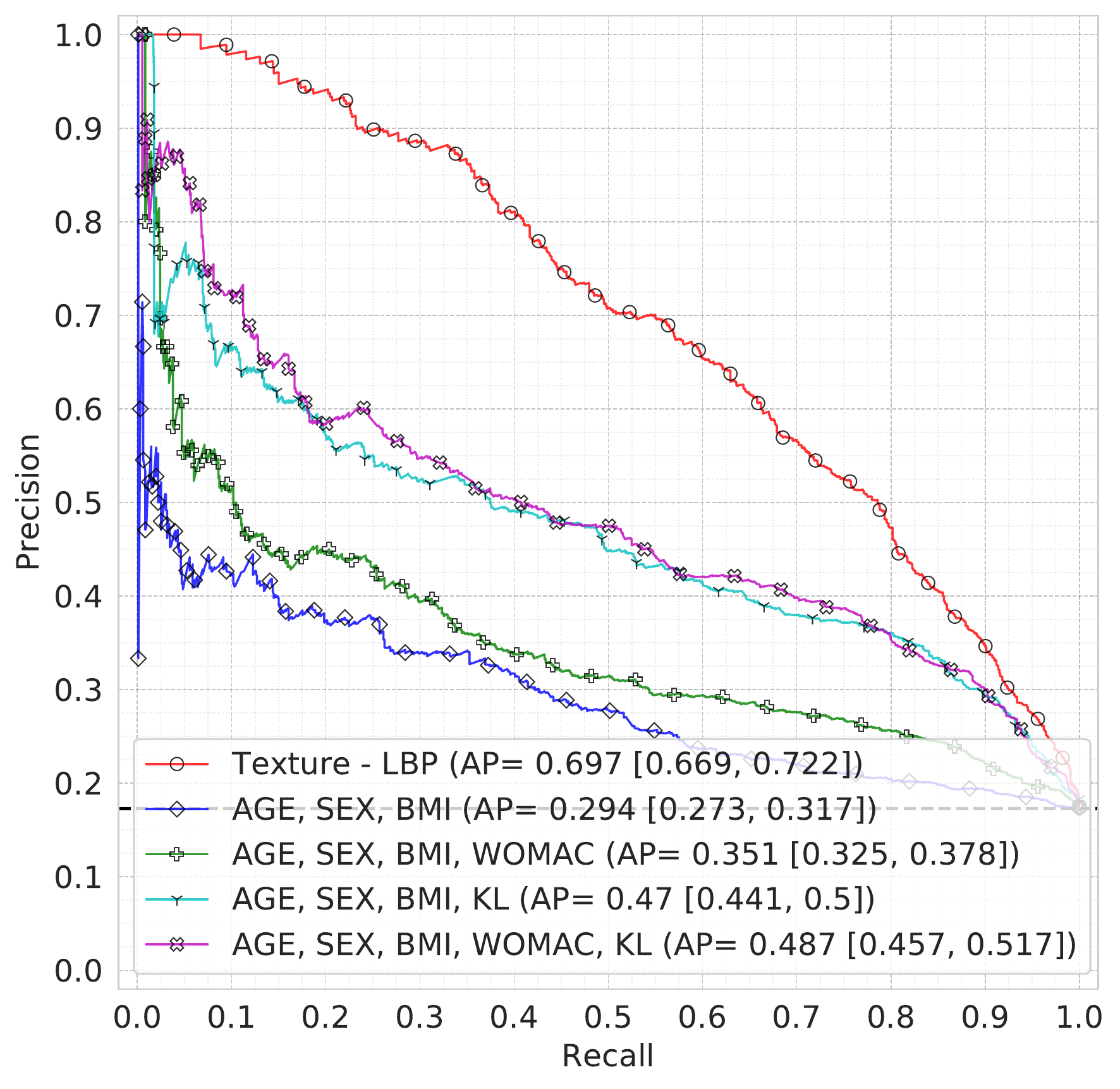}
    \hfill}
    \caption{ Comparison of texture features and clinical data. (a) ROC and (b) PR curves demonstrating the performance of the gradient boosting machine (GBM) models. 
    Superior region of interest(ROI) with Local Binary Pattern (LBP) features were employed to train the texture model.
    Area under the curves and 95\% confidence intervals in parentheses were given based on a 5-fold cross validation setting.
    Dashed lines in ROC indicate the performance of a random classifier and in case of PR it indicates the distributions of the labels of the dataset (PFOA vs non-PFOA). }
    \label{fig:features_comparison}
\end{figure}


\subsection*{{Comparison of Texture Features and Clinical Variables}}

After testing the ROIs, we developed machine learning models based on clinical assessments and subject characteristics.
In order to understand the value of texture features, we performed a thorough evaluation of age, sex, body-mass index (BMI), WOMAC and TFOA KL scores (Figure \ref{fig:features_comparison}).
Here, we utilised GBM models and trained them to predict the probability of PFOA using different combinations of the risk factors mentioned above. 
Figure \ref{fig:features_comparison} demonstrates that the best performed reference model is based on age, sex, body-mass index, WOMAC, and TFOA KL scores (Model5).
It reached the AUC of 0.817 (0.802–0.831) and AP of 0.487 (0.457–0.517).
When we compared our texture model (Model1) to the strongest reference method (Model5),
we obtained a statistically significant difference between the AUC values (DeLong's p-value$<1e-5$). 

\subsection*{{Comparison of Texture Features and Deep Convolutional Neural Network Features}}
Subsequently, we developed a CNN model, which directly works on texture patches in an automatic manner.
Compared to the handcrafted texture model (Model1), our CNN model(Model6) yielded slightly higher performance with an AUC of 0.889(0.875-0.9) and AP of 0.714(0.687-0.739), however the  performance difference in AUC was not statistically significant (DeLong's p-value=0.25) (Table \ref{tab:comparison}).
ROC and PR curves are shown in the Supplementary Material (Figure \ref{fig:texture_cnn_comparison}).

\subsection*{{Model Stacking and Early Feature Fusion}}

\begin{figure} [!htb]
    \centerline{
    \includegraphics[width=0.8\textwidth]{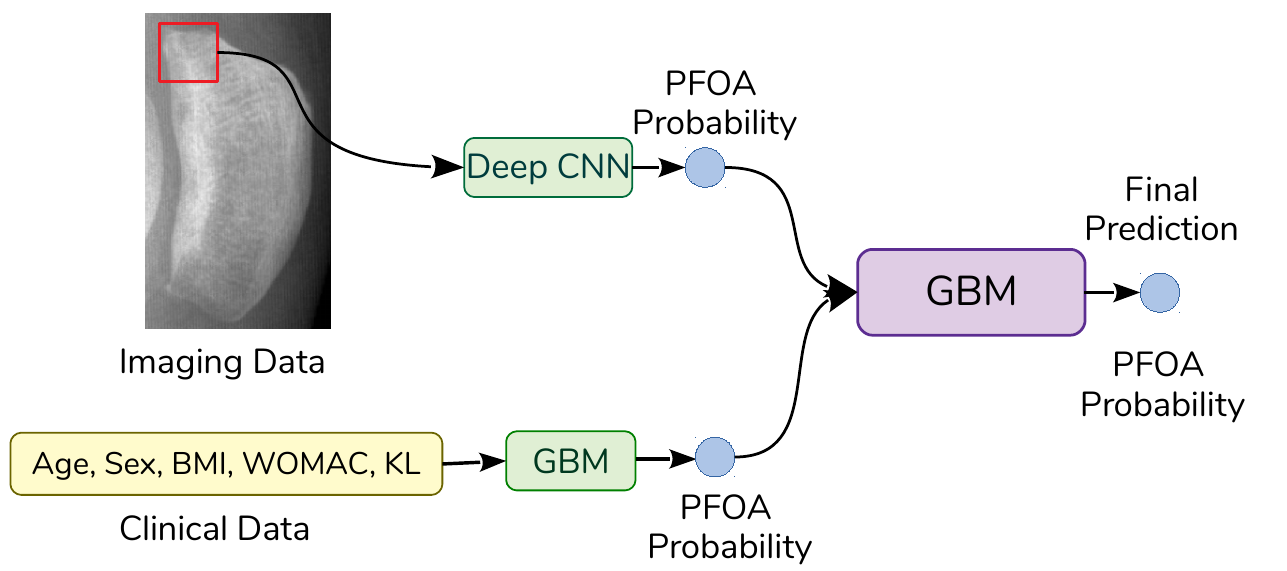}
    }
    \caption{ Schematic representation of model stacking.}
    \label{fig:model_stacking}
\end{figure}

Finally, in order to evaluate whether a combination of texture features and clinical assessments and participant characteristics would further increase the predictive accuracy, we utilised a second layer GBM model that fuses model predictions (model stacking). 
Here, we used predictions of our CNN model (Model6) and predictions of the strongest reference model (Model5)  as input features for the second level GBM model (Figure \ref{fig:model_stacking}). Same 5-fold stratified cross validation setup was used. 
This model (Model8) yielded the best AUC of 0.937 (0.929, 0.945) and AP of 0.791 (0.768, 0.813) and the Brier score of 0.069.
This increase in AUC was also statistically significant (DeLong's p-value$<1e-5$) between the stacked model (Model8) and the CNN model (Model6). 

For comprehensiveness, we also tested fusion of texture features (LBP) and clinical assessments and participant characteristics (age, sex, BMI, WOMAC, KL) in the same GBM model (See Supplementary Figure \ref{fig:fused_model}). 
ROC and PR curves of the fused model are shown in the Supplementary Material in Figure \ref{fig:fused_comparison}.
Fused model yielded AUC of 0.904 (0.892, 0.913) and AP of 0.719 (0.693, 0.743).
While fused model (Model7) performed significantly better than the texture model (Model1) in terms of AUC (DeLong's p-value$<1e-5$), it is still not better than the stacked model (Model8).

%% file: Conclusion.tex
\section*{Discussions and Conclusions}

In this study, we presented a machine learning method based on textural features of patella to detect PFOA from lateral view knee x-rays.  Texture features were extracted both by handcrafted descriptor (LBP) and also learned directly from texture patches using CNNs.  We compared-texture based models with more conventional models that use clinical assessments and participant characteristics (age, sex, BMI, WOMAC,KL). 
The model that uses only texture features obtained from superior ROI from patellar region yielded statistically significant improvement over the clinical features (ROC AUC of 0.884 vs. 0.817). This finding suggests that patellar texture features can provide useful imaging biomarkers for OA diagnostics.

The highest ROC AUC and AP for classifying knees without and with PFOA were obtained when model predictions based on texture features and clinical data were combined into a second level machine learning model. 
This can be explained by the fact that the textural features of patellar bone and the clinical features are complementing each other. 
Model stacking outperformed early feature fusion because the former method allow
each classifier to include their own benefits.

We tested three different ROIs for texture analysis. 
Texture features from the superior ROI had the highest classification performance to distinguish between knees with and without PFOA. 
We observed a performance decrease when we used the whole patellar region for predicting PFOA compared to superior and inferior ROI.
These findings suggest that the area closest to the inferior margin of the patella, where osteophytes typically occur, experience the most significant bony changes in PFOA. 

This study has also limitations.
The most important one is the lack of external data for validating the performances of the models. 
Therefore, we adopted the cross fold validation strategy to evaluate them. 
However, the actual generalizability of the models could only be understood on a separate test data that comes from different sources (population, hospital, device, etc.).
However, because the patellofemoral OA is not studied as much as tibiofemoral OA, the number of PFOA-focused dataset is limited.
Second, PFOA is not only altering the bone structure but at the same time changes the morphology and the alignment of patella \cite{macri2017patellofemoral}. 
In future studies, models that consider patella morphology and alignment should be explored and the additional value of textural features on the diagnosis of PFOA should be studied.
Third, other views of patella such as skyline view should be analysed to confirm the location of the bone alterations.
Finally, in order to extract texture features, we employed rectangular ROIs. In some cases such ROI could overlap with the background that might slightly affect the results. 
In principle, adopting ROIs that adhere to the boundaries would be better, but this requires exact segmentation of the patellar region, and it might also lead to a non-constant ROI size further affecting the standardized comparisons between different subjects.

In conclusion, we present the first study that evaluates patellar bone texture for detecting PFOA. Our results show that texture features of patellar bone are different between knees with and without PFOA. Good classification performance values indicate that analyzed texture features contain useful information of patellar bone structure, which seems to change in PFOA. These texture features may be used in the future as novel imaging biomarkers in OA diagnostics.

%% file: supplementary.tex
\subsection*{Deep Convolutional Neural Network Architecture and Training Strategy}

\begin{figure}[!ht]
\centering
\includegraphics[width = 0.5\linewidth, ]{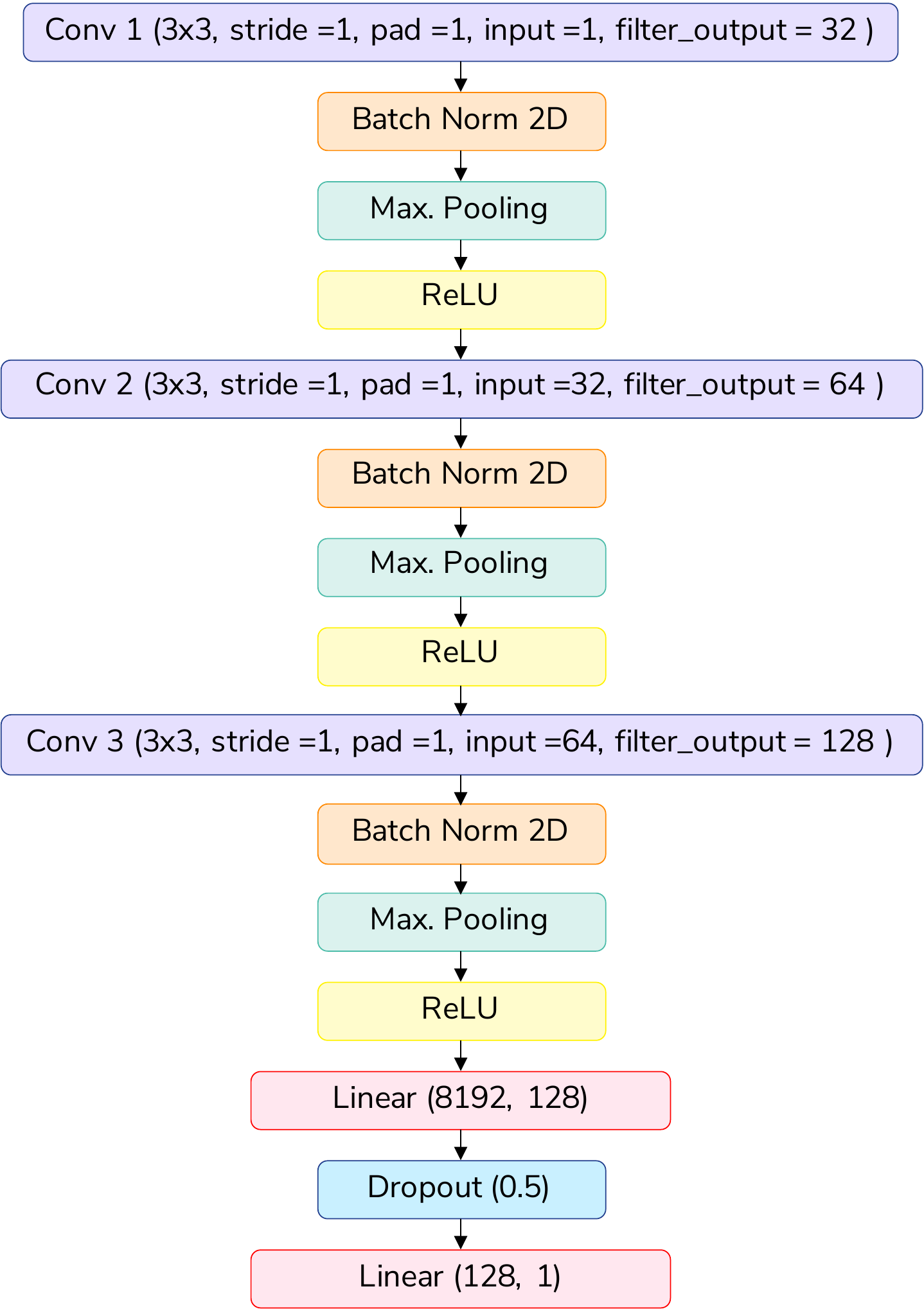}
\caption{Network Architecture }
\label{fig:network}
\end{figure}

The CNN model consists of 3 convolutional layers dedicated to texture feature extraction.
Each  convolution layer (stride=1, padding =1) is followed by Batch normalization (BN), max pooling ($2\times2$) and ReLU.
After feature extraction, we used two fully connected layers to make the prediction.
A dropout of 0.5 is inserted after the first fully connected layer.

In all the CNN based experiments, we used the same training strategy.
We trained the models from scratch (end-to-end) using the random weight initialization.
We adopted stochastic gradient descent training on a GPU. 
A mini-batch of 64 images were employed, and a momentum of 0.9 was used and trained without weight decay. 
We used a starting learning rate of 0.01 and decreased it by 10 every 8 epochs. 
The models were trained for 40 epochs and we selected the best performed model.

\begin{figure} [!h]
    \centerline{
    \includegraphics[width=0.6\textwidth]{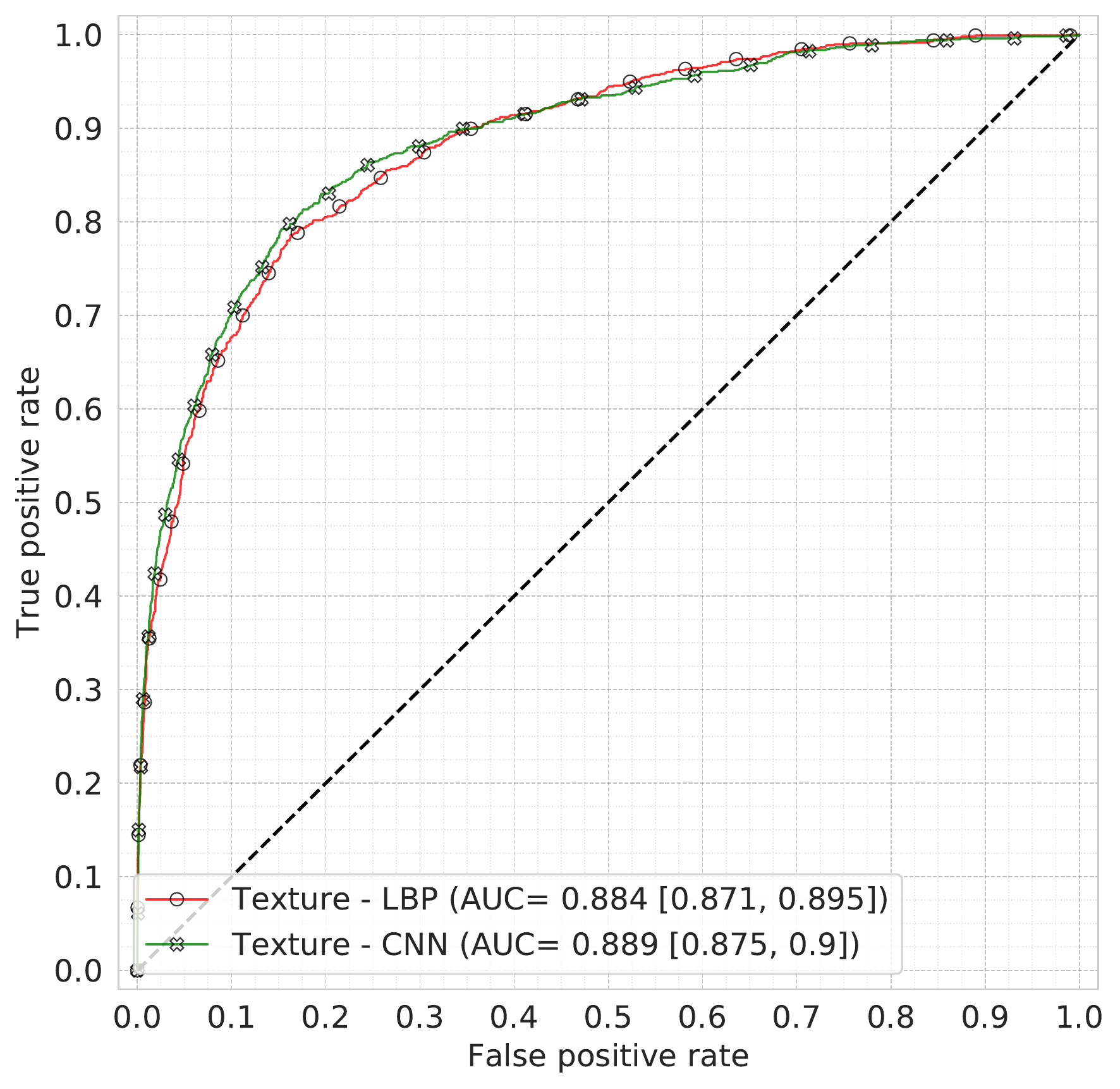}
    \includegraphics[width=0.6\textwidth]{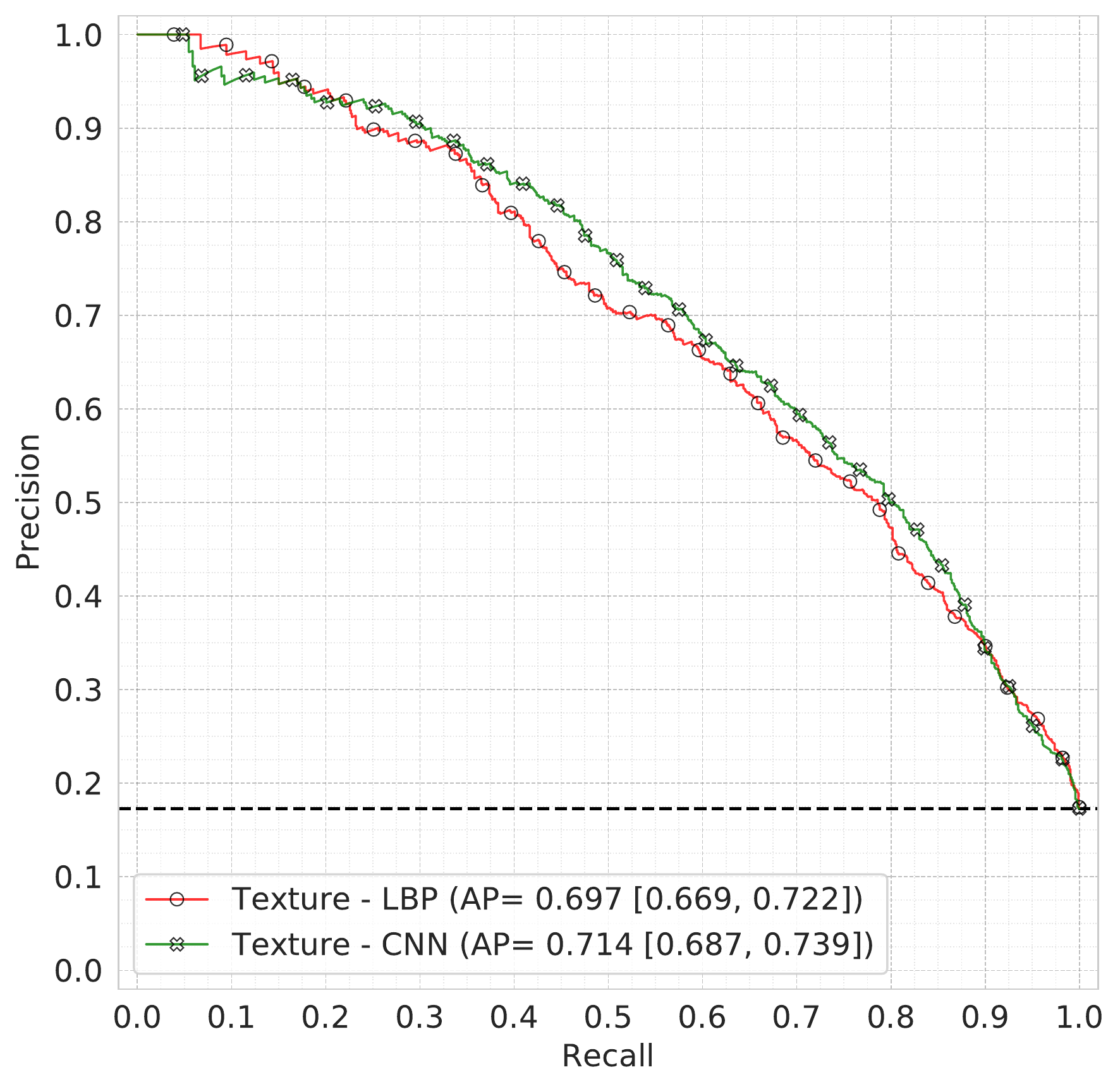}\hfill}
    \caption{ (a) ROC and (b) PR curves demonstrating the performance of the texture models- \textbf{CNN vs LBP}. 
    Area under the curves and 95\% confidence intervals in parentheses were given based on a 5-fold cross validation setting.
    Dashed lines in ROC indicate the performance of a random classifier and in case of PR it indicates the distributions of the labels of the dataset (PFOA vs non-PFOA).}
    \label{fig:texture_cnn_comparison}
\end{figure}


\begin{figure} [!htb]
    \centerline{
    \includegraphics[width=0.8\textwidth]{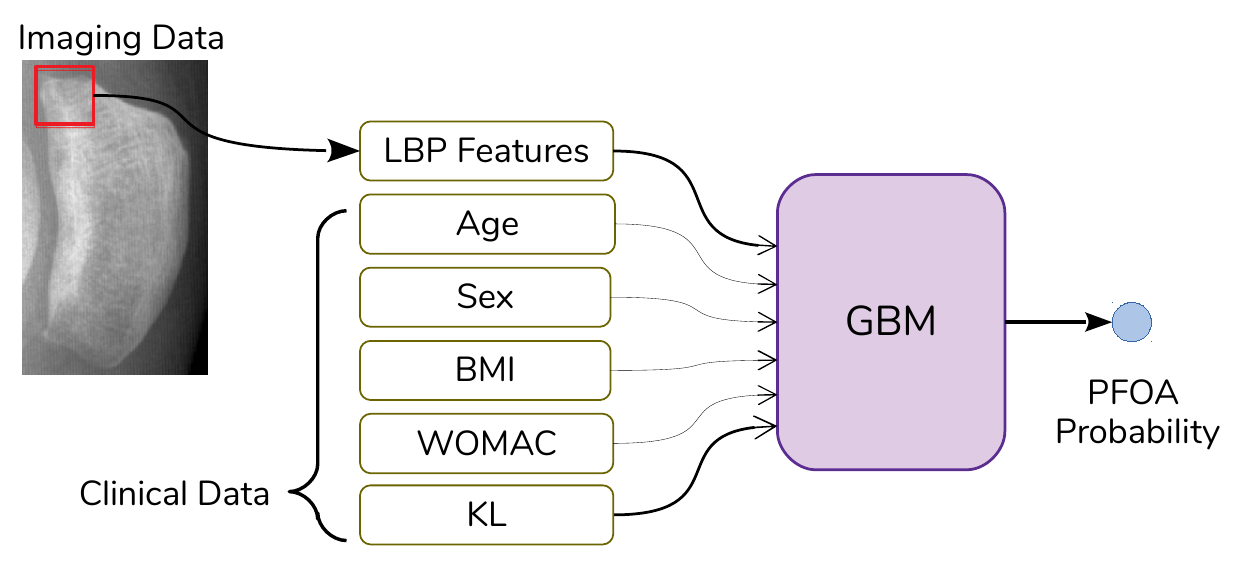}
    }
    \caption{ Schematic representation of feature fusion where texture features (LBP) and clinical data were combined in the same model. Performance of the fused model is shown in Figure \ref{fig:fused_comparison}. }
    \label{fig:fused_model}
\end{figure}

\begin{figure} [!htb]
    \centerline{
    \includegraphics[width=0.6\textwidth]{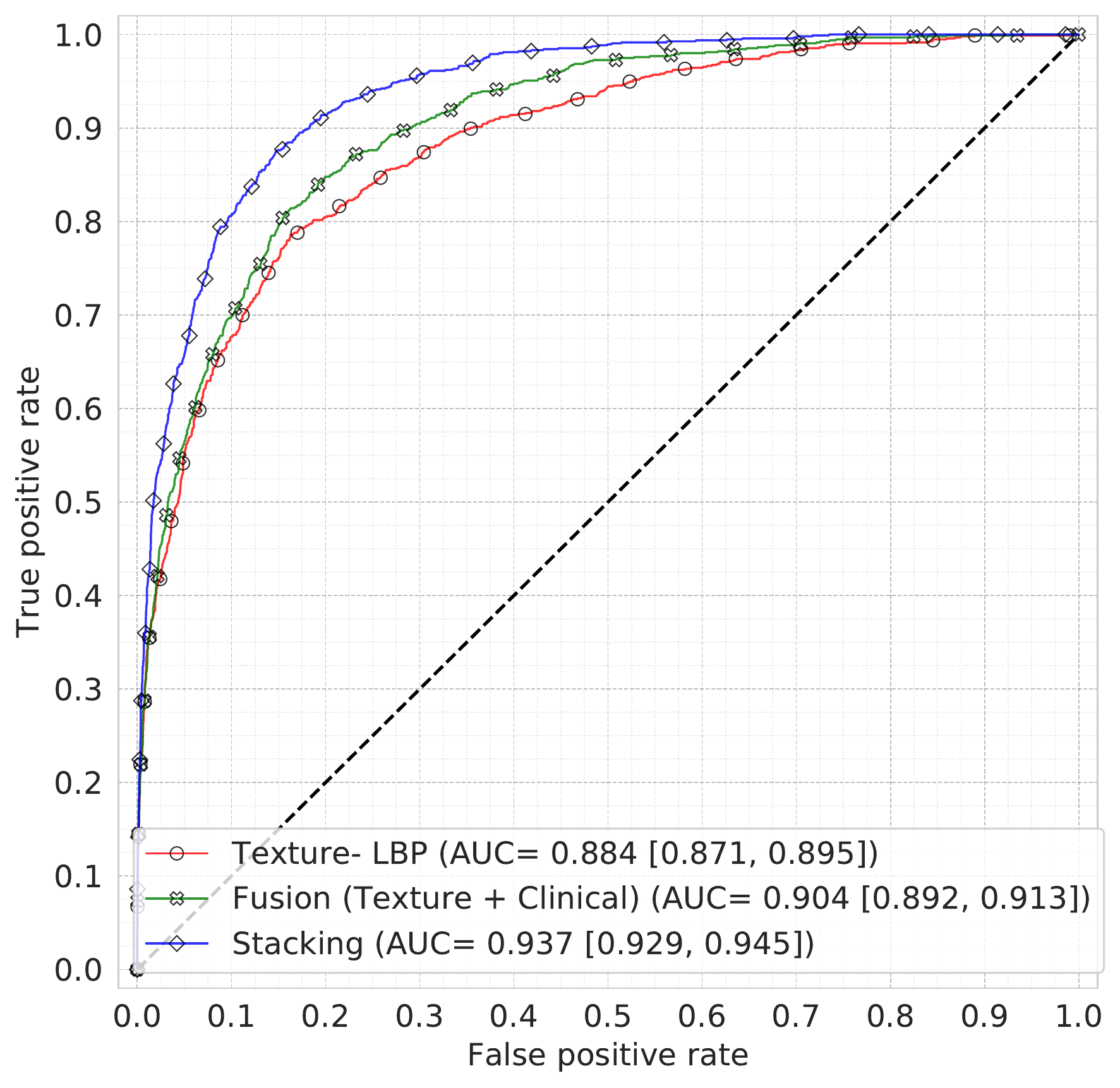}
    \includegraphics[width=0.6\textwidth]{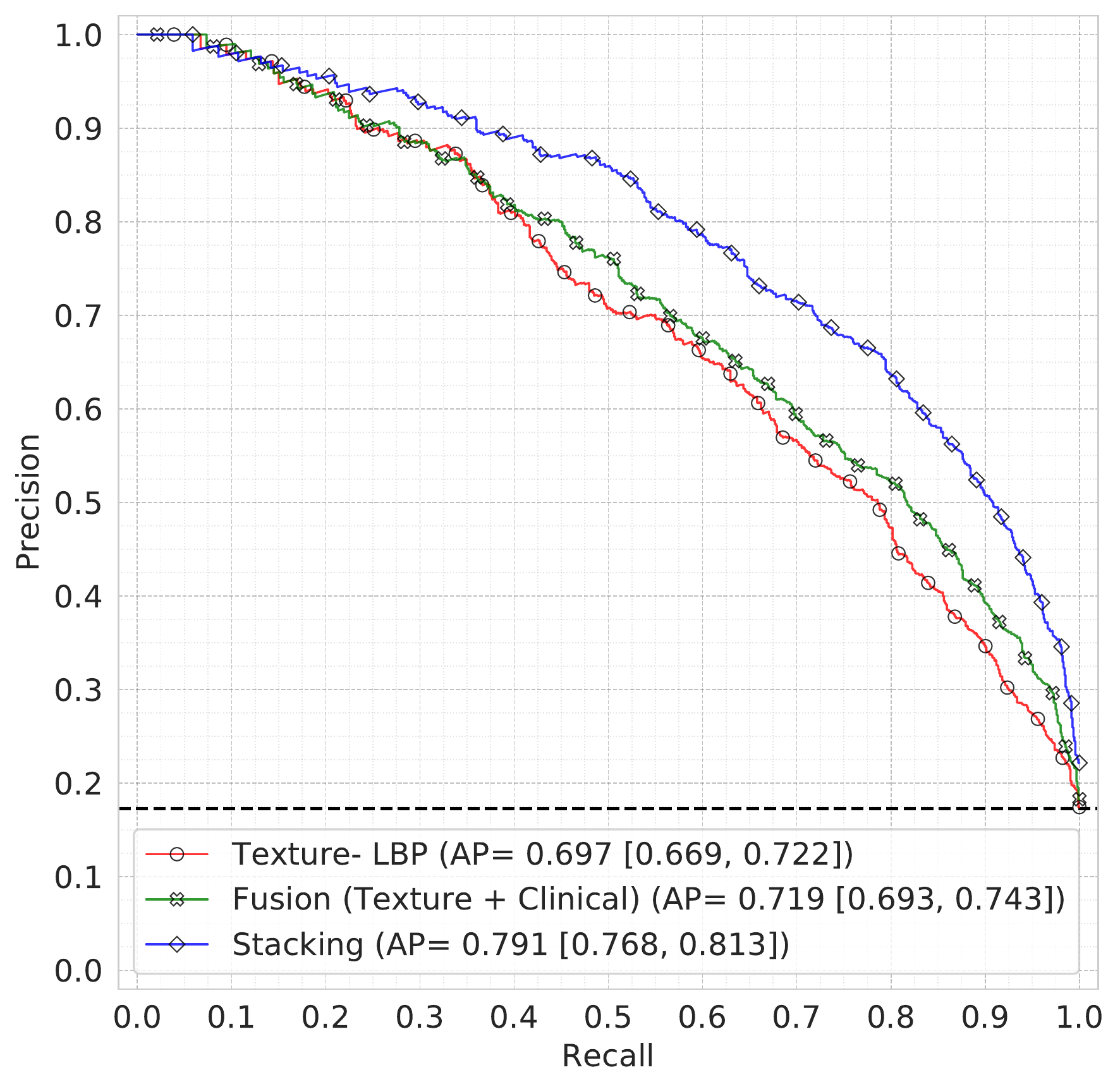}\hfill}
    \caption{ (a) ROC and (b) PR curves demonstrating the performance of the texture (LBP) model, fused model (that combines texture features and clinical data in the same model as shown in Figure \ref{fig:fused_model}) and stacked model (Figure \ref{fig:model_stacking}). 
    Area under the curves and 95\% confidence intervals in parentheses were given based on a 5-fold cross validation setting.
    Dashed lines in ROC indicate the performance of a random classifier and in case of PR it indicates the distributions of the labels of the dataset (PFOA vs non-PFOA).}
    \label{fig:fused_comparison}
\end{figure}